\documentclass[preprint,onecolumn,amsmath,amssymb,floatfix]{revtex4-1}
\usepackage{graphicx}
\usepackage{times}
\usepackage{color}

\begin{document}

\title{Vibrational and electronic heating in nanoscale junctions}

\author{Daniel R. Ward,$^{1}$ David A. Corley,$^{2}$, James M. Tour,$^{2}$, Douglas Natelson$^{1,3}$}

\affiliation{$^{1}$Department of Physics and Astronomy, Rice University, 6100 Main St., Houston, TX 77005, USA}
\affiliation{$^{2}$Department of Chemistry, Rice University, 6100 Main St., Houston, TX 77005, USA}
\affiliation{$^{3}$Department of Electrical and Computer Engineering, Rice University, 6100 Main St., Houston, TX 77005 USA}


\begin{abstract}
Understanding and controlling the flow of heat is a major challenge in nanoelectronics.   When a junction is driven out of equilibrium by light or the flow of electric charge, the vibrational and electronic degrees of freedom are, in general, no longer described by a single temperature\cite{Chen2003,Galperin2005,Dagosta2006,Pecchia2007,Huang2007,Galperin2007}.  Moreover, characterizing the steady-state vibrational and electronic distributions {\it in situ} is extremely challenging.  Here we show that surface-enhanced Raman emission may be used to determine the effective temperatures for both the vibrational modes and the flowing electrons in a biased metallic nanoscale junction decorated with molecules\cite{Ward2008}.  Molecular vibrations show mode-specific pumping by both optical excitation\cite{Galloway2009} and dc current\cite{Ioffe2008}, with effective temperatures exceeding several hundred Kelvin.  AntiStokes electronic Raman emission\cite{Moskovits2005,Otto2006} indicates electronic effective temperature also increase to as much as three times its no-current values at bias voltages of a few hundred mV.  While the precise effective temperatures are model-dependent, the trends as a function of bias conditions are robust, and allow direct comparisons with theories of nanoscale heating.  
\end{abstract}

\noindent Published as D. R. Ward, D. A. Corley, J. M. Tour, and D. Natelson, {\it Nature Nano.} {\bf 6}, 33-38 (2011), http://dx.doi.org/10.1038/nnano.2010.240

\maketitle

When a bias, $V$, is applied to a nanoscale junction, the electronic distribution is driven out of equilibrium, with ``hot" electrons injected from the source electrode into empty states $eV$ above the Fermi level of the drain electrode.  The electronic distribution rethermalizes on a scale set by inelastic electron-electron scattering.  The Landauer-B\"uttiker approach assumes that dissipative processes are spatially removed from the junction region, so that the source and drain electronic distributions have identical temperatures\cite{Datta1995}.   On longer time and distance scales, electron-phonon scattering transfers energy to the lattice.  If some current flows through molecules, inelastic processes can also transfer energy from the electrons to local molecular vibrational modes\cite{Stipe1998,Park2000}.  In the presence of optical radiation, the situation is even more complicated\cite{Galperin2009}.  Direct optical absorption and the decay of plasmon excitations produce hot electrons and holes in the metal with energies as large as those of the incident photons.  Raman scattering processes, hot electrons, and a hot substrate can also excite molecular vibrational modes, even in the absence of direct bias-driven processes\cite{Galloway2009}.

Directly accessing the electronic and vibrational distributions in a nanoscale junction is extremely difficult.  Like others before us,\cite{Dagosta2006,Pecchia2007,Huang2007,Ioffe2008} we assume that one may parametrize these (generally complicated) distributions by effective temperatures.  Previous experiments have used the stability of mechanical break junctions as a proxy for an effective ionic temperature\cite{Smit2004,Huang2007,Tsutsui2008}.

Raman spectroscopy is a more
direct probe of energy distributions, and the effective temperature of a particular vibrational mode, $T_{\nu}^{\mathrm{eff}}$, can be defined as\cite{OronCarl2008,Ioffe2008,Berciaud2010}  
\begin{equation}
\frac{I_{\nu}^{\mathrm{AS}}}{I_{\nu}^{\mathrm{S}}}=A_\nu \frac{(\omega_{\mathrm{L}}+\omega_{\nu})^4}{(\omega_{\mathrm{L}}-\omega_{\nu})^4} \exp(-\hbar \omega_\nu / k_{\mathrm{B}} T_\nu^{\mathrm{eff}}),
\end{equation}
where $I_{\nu}^{\mathrm{S}}$ and $I_{\nu}^{\mathrm{AS}}$ are the Stokes and antiStokes Raman intensities for that mode, $\omega_{\mathrm{L}}$ is the incident laser frequency, $\hbar \omega_{\nu}$ is the Raman shift, and $A_{\nu}$ is a correction factor that accounts for the ratio of the antiStokes and Stokes cross-sections.  In surface-enhanced Raman spectroscopy (SERS), plasmons enhance Raman response by orders of magnitude\cite{Moskovits2005}.  In this case, $A_{\nu}$ also accounts for differences in the Raman enhancement between the antiStokes and Stokes frequencies.  We assume that $A_{\nu}$ does not vary significantly with bias, since plasmonic properties are determined by metal geometry and dielectric function.  Bias-driven changes in the intrinsic antiStokes/Stokes crossections would require Stark-like physics, and would depend strongly on molecular orientation and mode.

Optical pumping may also elevate $T_{\nu}^{\mathrm{eff}}$ above the ambient temperature, $T$.  If the Stokes process occurs faster than vibrational relaxation to the bulk substrate, the excited vibrational population will exceed the equilibrium average.  Moreover, optically excited energetic electrons in the metal can couple to molecular vibrations.  In this regime, $I_{\nu}^{\mathrm{AS}}$ would scale quadratically with incident laser intensity, while $I_{\nu}^{\mathrm{S}}$ scales linearly as usual\cite{Galloway2009}.
  
Nanojunction devices open the possibility of observing bias-driven vibrational pumping via antiStokes/Stokes SERS measurements, as suggested theoretically\cite{Galperin2009} and experimentally\cite{OronCarl2008,Ioffe2008,Berciaud2010}.  Cross-sections for electron-vibrational pumping are expected to differ from mode to mode, as are the relaxation rates to the bulk substrate.  Conversely, intra-molecular vibrational relaxation would push all vibrational modes toward a single effective temperature.

Finally, in SERS systems with strong enhancements, a continuum of Raman scattering is observed from the conduction electrons in the metal\cite{Zawadowski1990,Jiang2003,Moskovits2005,Otto2006,Mahajan2010}.  Stokes scattering results from creation of electron-hole excitations, while antiStokes scattering requires excited electrons above the Fermi level of the metal.  Crystal momentum kinematically limits the accessible Raman shifts, with disorder and the surface playing critical roles\cite{Zawadowski1990,Otto2006}.  There is no complete theoretical description of the full electronic continuum\cite{Mahajan2010}.  However, the antiStokes Raman response at some shift $\epsilon$ is expected\cite{Otto2006} to be proportional to the joint density of states (JDOS) for electrons at energy $\epsilon+E$ and holes at energy $E$.  Taking the kinematic restriction and the band densities of states to be slowly varying over the antiStokes shift range of interest, and neglecting any resonance effects, we expect
\begin{equation}
I_{e}^{\mathrm{AS}}(\epsilon) \propto \int{f(E+\epsilon,T_{e}^{\mathrm{eff}})(1-f(E,T_{e}^{\mathrm{eff}}))dE}=\frac{\epsilon}{e^{\epsilon/k_{\mathrm{B}}T_{e}^{\mathrm{eff}}}-1},
\end{equation}
where the integral is over the conduction band, $f(E,T)$ is the Fermi-Dirac function, and $T_{e}^{\mathrm{eff}}$ is an effective temperature for the conduction electrons.  Only Raman emission from the electrons that are proximate to the SERS ``hotspot'' is readily detected, since only that emission is appropriately enhanced.  

We perform Raman spectroscopy on pairs of gold nanojunction electrodes, with the fabrication and measurement processes described in Methods.  Prior to electromigration\cite{Park1999} a monolayer of the molecule of interest is self-assembled on the electrodes.  Two molecules were examined:  a three-ring oligophenylene vinylene terminated in amine functional groups (OPV3, described further in Supplementary Information) and 1-dodecanethiol.  Electromigration and measurements take place in vacuum with the substrate at 80~K.  Such nanogaps are excellent SERS substrates\cite{Ward2007}, and have a long track record in tunnelling transport through single molecules\cite{Park2000,Natelson2006}.  In nanojunctions with measurable conductances, prior work\cite{Ward2008,Ward2008b} and the present experiments (see Supplementary Information) demonstrate strong correlations in the time variation of the nanojunction conductance and ``blinking'' and spectral diffusion of the SERS spectrum.  The tunnelling conductance probes a very small volume.  Hence, correlations between the tunnelling conductance and the SERS signal imply single- or few-molecule SERS sensitivity.  Final nanojunction conductances range from 0.05~$G_{0}$ to 0.01~$G_{0}$ (where $G_{0}\equiv 2e^2/h$ is the quantum of conductance) for junctions with molecules used in heating measurements, higher than that expected for an idealized gap bridged by such a molecule attached to both electrodes\cite{Venkataraman2006}.  For measurements of electronic temperature, nanojunction conductances ranged from 0.5~$G_{0}$ to 0.01~$G_{0}$, with the highest conductance nanojunctions showing no signs of vibrational Raman indicating the absence of molecules in the nanojunction.  For lower values of the conductance, direct metal-metal tunnelling continues to be responsible for much of the total conductance, but the temporal correlations between conductance and Raman spectra imply that at least some of the tunnelling current interacts with the molecule (or molecules) even if it (they) does not neatly bridge the gap between the electrodes (see Supplementary Information).  An image of a typical nanojunction is shown in Fig.~1a, and the measurement scheme is shown in Fig.~1b.

Simultaneous optical and electrical measurements provide a wealth of information, as seen in Figure 1c, d.  For this device we can observe optical and electrical pumping of individual vibrations of OPV3.  We also observe apparent heating of the electrons as evidenced by the rise in low wavenumber antiStokes continuum with increasing magnitude of bias voltage.  For simplicity in presentation, we discuss the effective temperature analysis in terms of optical vibrational pumping, electrical vibrational pumping, and electron heating in turn.  

%
%
Optical pumping of vibrations ($T_{\nu}^{\mathrm{eff}} > T$ at $V_{DC}=0$) was observed in nanojunctions for both test molecules.  Figure 2a shows the evolution of the Raman response of a dodecanethiol junction as a function of time.  Several different discrete spectral configurations are observed, as seen on the Stokes side of the spectrum.  For each of these configurations, we observe the same two modes (1099~cm$^{-1}$ and 1475~cm$^{-1}$) in the antiStokes spectrum, but with varying intensities.  Effective temperatures are inferred using Eq.~(1), assuming the Raman enhancement is frequency independent and the cross-sections for Stokes and antiStokes scattering are equal ($A_\nu = 1$).  Effective temperatures range from 300~K to greater than 2000~K, demonstrating that vibrations may be driven far from equilibrium via optical excitation alone.  Note that when the spectrum returns to a previously visited configuration, the effective temperatures also return to the previous values associated with that spectrum.  The simplest explanation for the simultaneous changes in spectrum and effective temperatures is that the molecule(s) in question undergo conformational changes relative to the Au electrodes.  Such changes in conformation can affect SERS enhancement, vibrational relaxation times, electronic coupling between the molecule and electrodes, and possibly spectral diffusion as well.  Any of these changes could result in different pumping rates.  Note that the two modes presented here are not in equilibrium with one another, with an effective temperature difference of more than 300~K between them.  As shown in Fig. 2c, we observe a quadratic dependence of the antiStokes signal on incident power, as expected in the optical pumping regime.\cite{Galloway2009}

Under DC bias we observe strong electrical pumping of specific vibrational modes in both OPV3 and dodecanethiol.  Additional examples are given in Supplementary Information.  Figures 3a and d show the bias dependence of $T_{\nu}^{\mathrm{eff}}$ for different modes for two different devices with OPV3 assembled on them.  We observe an approximately linear increase in $T_{\nu}^{\mathrm{eff}}$ with increasing $|V|$.  This observation across multiple modes, multiple devices, and multiple molecules makes it very unlikely that this trend results from some bias-driven change in $A_{\nu}$.  For the sample in Figure 3a-c we are signal limited below $V = 0.2$~V and observe no measureable antiStokes signal above our detector noise floor.  There is a voltage asymmetry in conductance and $T_{\nu}^{\mathrm{eff}}$ between the positive and negative sweeps.  On the positive $V$ side, the 1625~cm$^{-1}$ mode is only slightly higher in effective temperature than the 1317~cm$^{-1}$ mode, while on the negative $V$ side, the difference is considerably larger.  This dichotomy is reproduced on multiple voltage sweeps on this device.  One possible explanation for this sort of mode-to-mode variation and voltage asymmetry is that the molecule's electronic coupling is different for the two electrodes, so the effective cross-sections for electron-vibrational pumping depend on the direction of current flow.    Such an explanation has been suggested for asymmetries observed in vibration-mediated inelastic electron tunneling spectroscopy\cite{Galperin2005b}.   Figure 3b also shows another feature that we observe with some regularity in such junctions:   a systematic shift in energy of about 15~cm$^{-1}$ is observed for most (Stokes) vibrational modes as a function of $V$.  This is an example of a Raman Stark shift\cite{Lambert1984}.  The decrease in overall Raman intensity observed on this particular positive bias sweep is attributed this to slow drift of the laser spot, and does not appear to affect the effective vibrational temperature since we are in a regime with no detectable optical pumping.

We think it is unlikely that the vibrational heating results from heat flow from hot electrodes rather than the flow of charge through the molecule(s), for the following reason.  If the electrodes acted as an effective high temperature bath, all molecular modes with any coupling to that bath should reach a similar steady state temperature.  Instead, we observe only a few modes, and those have disparate temperatures and heating rates.  While plasmon dispersion in $A_{\nu}$ is a concern, numerical modeling of such junctions\cite{Ward2007} shows trends (stronger resonances to the red; resonances hundreds of meV broad in energy) inconsistent with that as a dominant effect.   While we cannot rule out a contribution due to energy flow from the electrodes to the molecule(s), we believe pumping of vibrations due to current\cite{Galperin2005b} is more consistent with these observations. 

For the device shown in Figure 3d, e, and f, both electrical and optical vibrational pumping are observed.  The clear 1815~cm$^{-1}$ antiStokes mode has measurable antiStokes signal at zero bias, with a temperature more than 300~K above the substrate temperature.  Until $V$ exceeds $\hbar\omega_\nu$, no increase in $T_{\nu}^{\mathrm{eff}}$ is observed, as expected.  Once $V$ exceeds this threshold, once again an approximately linear increase in $T_{\nu}^{\mathrm{eff}}$  is observed with increasing $V$.  These observations are consistent with theoretical treatments of electron-vibrational scattering during off-resonant conduction\cite{Pecchia2007,Galperin2008}.  We do not observe any signs of vibrational cooling at high biases\cite{Dagosta2006,Ioffe2008}; nor do we see $T_{\nu}^{\mathrm{eff}} \sim \sqrt{V}$ at lower biases, as would be expected for cooling of the local modes via relaxation to bulk 3d phonons\cite{Dagosta2006}.  This suggests that local vibrational relaxation to the substrate takes place via other means, such as coupling to phonons of reduced dimensionality.  We again observe some asymmetry in conductance (Figure 1d) and $T_{\nu}^{\mathrm{eff}}$ between the positive and negative sweeps.  The vibrations at 1480~cm$^{-1}$ and 562~cm$^{-1}$ show a strong asymmetry similar to the conductance, while the 1815~cm$^{-1}$ mode shows very little asymmetry.  This again suggests that effective cross-sections for electron-vibrational pumping differ from mode to mode.

In Figure 4 we examine heating of the electrons as a function of $V$.  Effective temperatures are found by fitting the antiStokes continuum spectra to Equation 2.  At $V=V_{\mathrm{max}}$, two fit parameters are used, $T_{e}^{\mathrm{eff}}$ and an overall amplitude.  At other bias voltages, the amplitude is then fixed, leaving $T_{e}^{\mathrm{eff}}$ as the only free parameter to fit at each $V$.  Representative spectra and fits are shown in Figs. 4c, f.    Error estimates for $T_{e}^{\mathrm{eff}}$  correspond to a doubling of $\chi^2$ from the best fit.    In Figure 4a we see a roughly 150~K change in temperature over a 400~mV change in bias.  There is asymmetry in both the temperature and current as a function of $V$.  One possible explanation for this would be that only one electrode gets hot at a time, depending on the direction of current flow, and the thermal coupling to the bulk is different for each side of the nanogap.  An increase of $T_{e}^{\mathrm{eff}}$ roughly linear in $V$ is consistent with a general argument based on steady state heat flow in the electron system\cite{Dagosta2006}.  

The observation of effective electronic heating with bias in nanoscale junctions is important.  While it is expected on general grounds\cite{Dagosta2006}, the Landauer-B{\"u}ttiker approach to conduction, commonly used to model such junctions, does not account for such an effect.  Additional experimental consequences would be expected, including broadening of inelastic electron tunneling spectra at high biases under similar current densities, and signatures of heating in the junction current noise.  Investigations in these directions are ongoing.

There are \emph{systematic} uncertainties in the effective vibrational and electronic temperatures (due to the unknown values of $A_{\nu}$ and the lack of a detailed theory of the electronic Raman continuum).  However, the \emph{trends} captured in the data (the dependence of the effective temperatures on the bias conditions, for example) are robust.  These experiments demonstrate that it is possible to extract highly local information about the effective ionic (vibrational) and electronic temperatures within nanoscale junctions under electrical bias.  The resulting data are among the first direct measurements that will allow specific tests of theoretical treatments\cite{Chen2003,Dagosta2006,Pecchia2007,Galperin2007} of thermal dissipation at the nanoscale.  It is clear that there can be great variation in electron-vibrational couplings and vibration-bulk relaxation rates; with weak relaxation to the bulk, electrical pumping could access vibrational temperatures comparable to $eV/k_{\mathrm{B}}$.  These experiments may open the door to deliberate control and engineering of such effects.

\noindent{\bf Methods}

Samples were fabricated on N-type Si with 200~nm of thermal oxide.  Nanowires approximately 120~nm wide were defined by e-beam lithography and e-beam evaporation of 0.5~nm Ti and 15~nm of Au.  The nanowires were approximately 600~nm long and connected to larger electrical leads and contact pads.  After fabrication samples were O$_2$ plasma cleaned and soaked in either 0.1~mg/mL OPV3/THF or 10~mM 1-dodecanethiol in ethanol for 24~hours to form a self-assembled monolayer on the Au surface.  After assembly samples were wire bonded to a ceramic chip carrier and placed in an optical cryostat.  Samples were then cooled to 80~K in vacuum.

Nanogaps were formed in the nanowires via the electromigration process which has been studied extensively.  Gap formation was performed with an automated computer program which stopped the electromigration process once the electrical conduction dropped below $2e^2/h \equiv G_0$ - the quanta of conductance. 

After electromigration optical measurements were performed by illuminating the sample surface at 785~nm with a Gaussian laser beam with a spot size of 1.9~$\mu$m, incident intensity of 23~kW/cm$^2$, and polarization aligned along the nanowire.  The laser beam was rastered over the sample and a spatial map of the Raman response of the Si substrate was measured at 520~cm$^{-1}$ as described in Ref.~\cite{Ward2008}.  The Au electrodes of the nanogap attenuate the silicon signal allowing precise positioning of the laser over the nanogap.  The SERS signal of the nanogap was then measured.  The SERS signal may also be spatial mapped to confirm that it originates from the nanogap.  The spectra were all corrected to adjust for the spectral sensitivity of the system over the measured wavelength range, by calibrating to a known blackbody source ($T$=2960~K).  Integration times for SERS spectra during voltage sweeps ranged from 5~s to 10~s depending on signal strength.

Electrical measurements of the nanogap were performed using a current amplifier and lock-in amplifier as illustrated in Figure 1b.  The lock-in amplifier applied a 10~mV AC  signal to the sample and measures the first harmonic response ($\propto dI/dV$).  A summing amplifier is used to add a DC signal ($V$) to the lock-in signal allowing the current amplifier to measure $I$.  In this configuration both $I$ and $dI/dV$ can be measured simultaneously as functions of $V$. 

To avoid concerns about thermal broadening or extrinsic (detector-limited) broadening of Raman peaks, the effective temperature analyses were performed based on absolute peak height above baseline rather than integrated intensity over some wavenumber range.  Error estimates take into consideration the fact that the peak heights have a $\sqrt{N}$ noise, where $N$ is the number of CCD counts over baseline.

\noindent{\bf[Supplementary Information]} accompanies this paper at www.nature.com/naturenanotechnology.  Reprints and permission information is available online at http://npg.nature.com/reprintsandpermissions/.

\noindent{\bf[Acknowledgments]}
D.N. and D.R.W. acknowledge support by Robert A. Welch Foundation grant C-1636 and the  Lockheed Martin Advanced Nanotechnology Center of Excellence at Rice (LANCER).  D.N. and D.R.W. acknowledge valuable conversations with M. Di Ventra, M. A. Ratner, and A. Nitzan.

\noindent{\bf[Author Contributions]}
D.R.W. fabricated the devices, performed all measurements, and analyzed the data.  D.N. supervised and provided continuous guidance for the experiments and the analysis. D.A.C. synthesized the OPV3 molecules under the supervision of J.M.T.  The bulk of the paper was written by D.R.W. and D.N.  All authors discussed the results and contributed to manuscript revision.

\noindent{\bf[Competing Interests]}  The authors declare that they have no competing financial interests.

\noindent{\bf[Correspondence]}  Correspondence and requests for materials should be addressed to D.N.~(email: natelson@rice.edu).

\clearpage

\begin{figure}
\begin{center}
\includegraphics[width=7in]{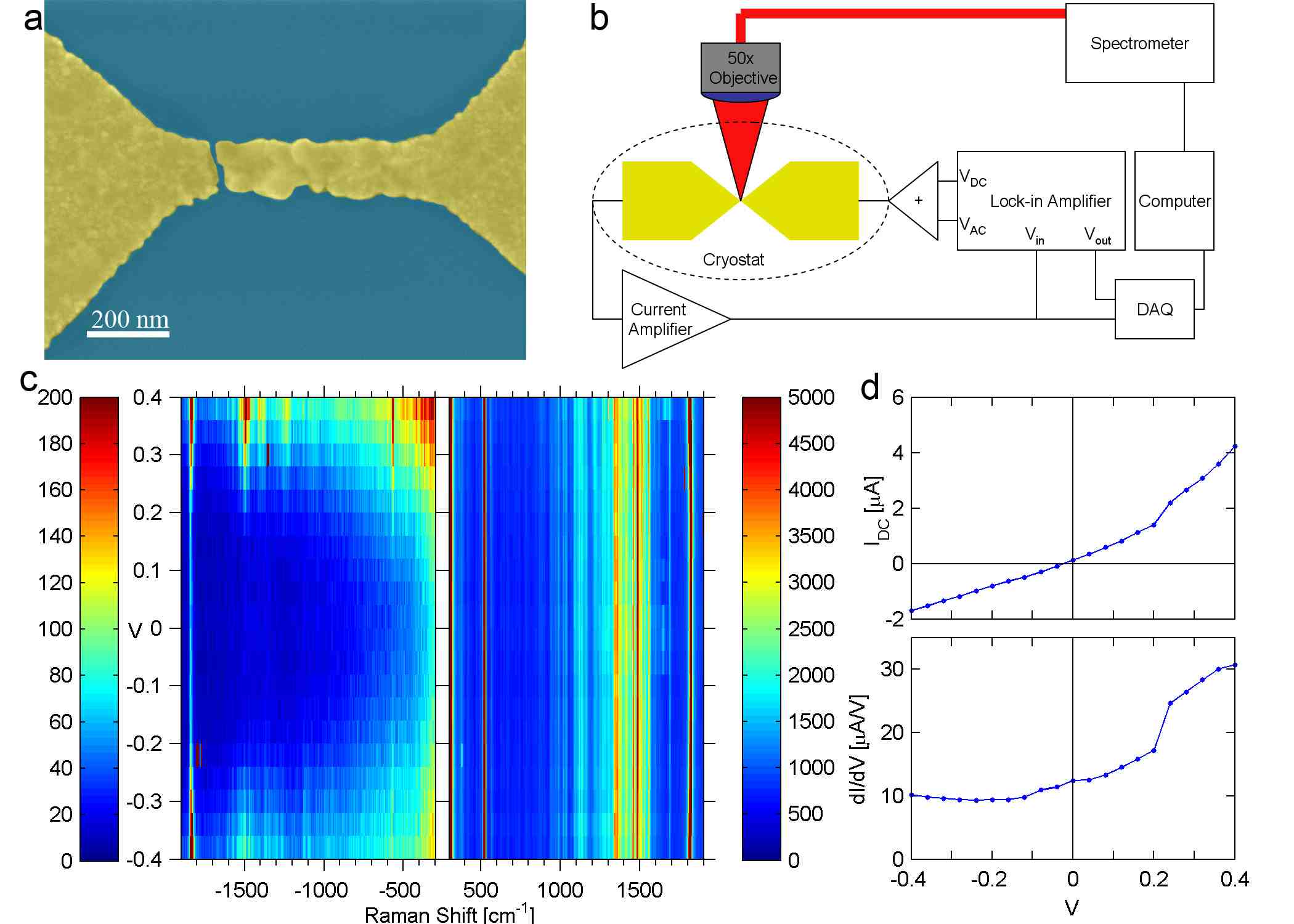}
\end{center}
\caption{{\bf Measurement overview.}
a) SEM false color image of a typical nanogap.
b) Schematic of electrical and optical measurement.
c) Waterfall plot of Raman response (in CCD counts) of an OPV3 junction as a function of DC bias, $V$.  The antiStokes spectrum shows strong dependence on $V$ while the Stokes side is relatively constant as a function of bias.  The strong Stokes peak at 520~cm$^{-1}$ is from the silicon substrate.
d) Simultaneously measured current and differential conductance as a function of $V$.  
}
\end{figure}

\clearpage

\begin{figure}
\begin{center}
\includegraphics[width=7.2in]{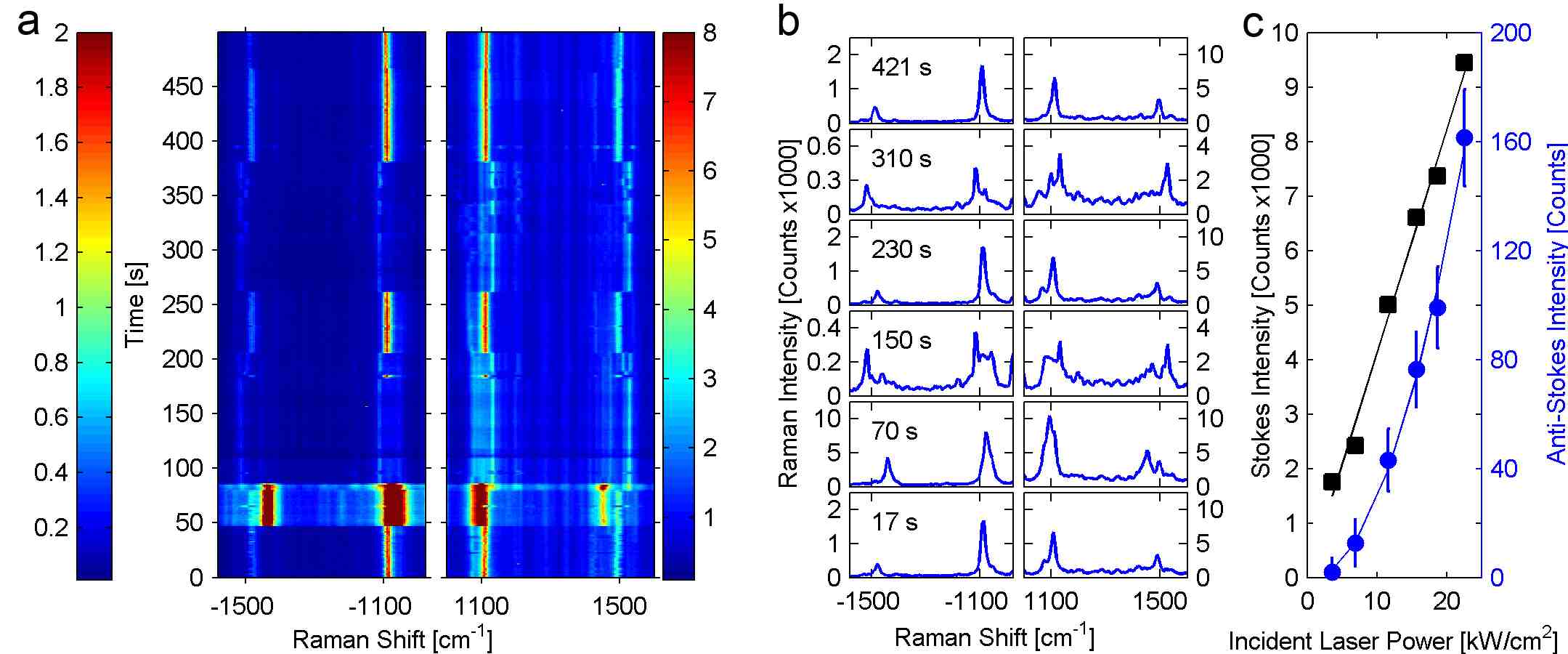}
\end{center}
\caption{{\bf Optically driven vibrational pumping.}
a) Raman response of a dodecanethiol-coated junction as a function of time under zero bias.  Scale bars are in thousands of CCD counts.  The junction switches stochastically between several stable configurations, each with characteristic spectra that exhibit strong optical pumping of vibrational modes at 1099~cm$^{-1}$ (C-C stretch) and 1475~cm$^{-1}$ (CH$_2$ or CH$_3$ deform). 
b) Spectra from given time slices in (a).  We note that the higher energy peak spectrally diffuses between  1475~cm$^{-1}$ and 1525~cm$^{-1}$ and the lower energy peak between 1099~cm$^{-1}$ and 1110~cm$^{-1}$.  The temperature of the 1099~cm$^{-1}$ mode in Kelvin is 769, 1722, 532, 754, 512, 775 with increasing time.  The temperature of the 1475~cm$^{-1}$ mode in is 701, 815, 658, 729, 648, 746.  We note that at 70~s a vibration at 1450~cm$^{-1}$ is observed at 2161~K.  All temperatures have an uncertainty of $\pm$ 10~K.
c) Power dependence of Stokes (black) and antiStokes (blue) signal taken from a different device also showing optical pumping.  As expected, in the optical pumping regime the Stokes signal is linear (solid line) in laser power while the antiStokes signal increases quadratically (solid line) in laser power.  Error bars indicate uncertainty in signal due to read and shot noise in CCD.  
}
\end{figure}

\clearpage

\begin{figure}
\begin{center}
\includegraphics[width=6.5in]{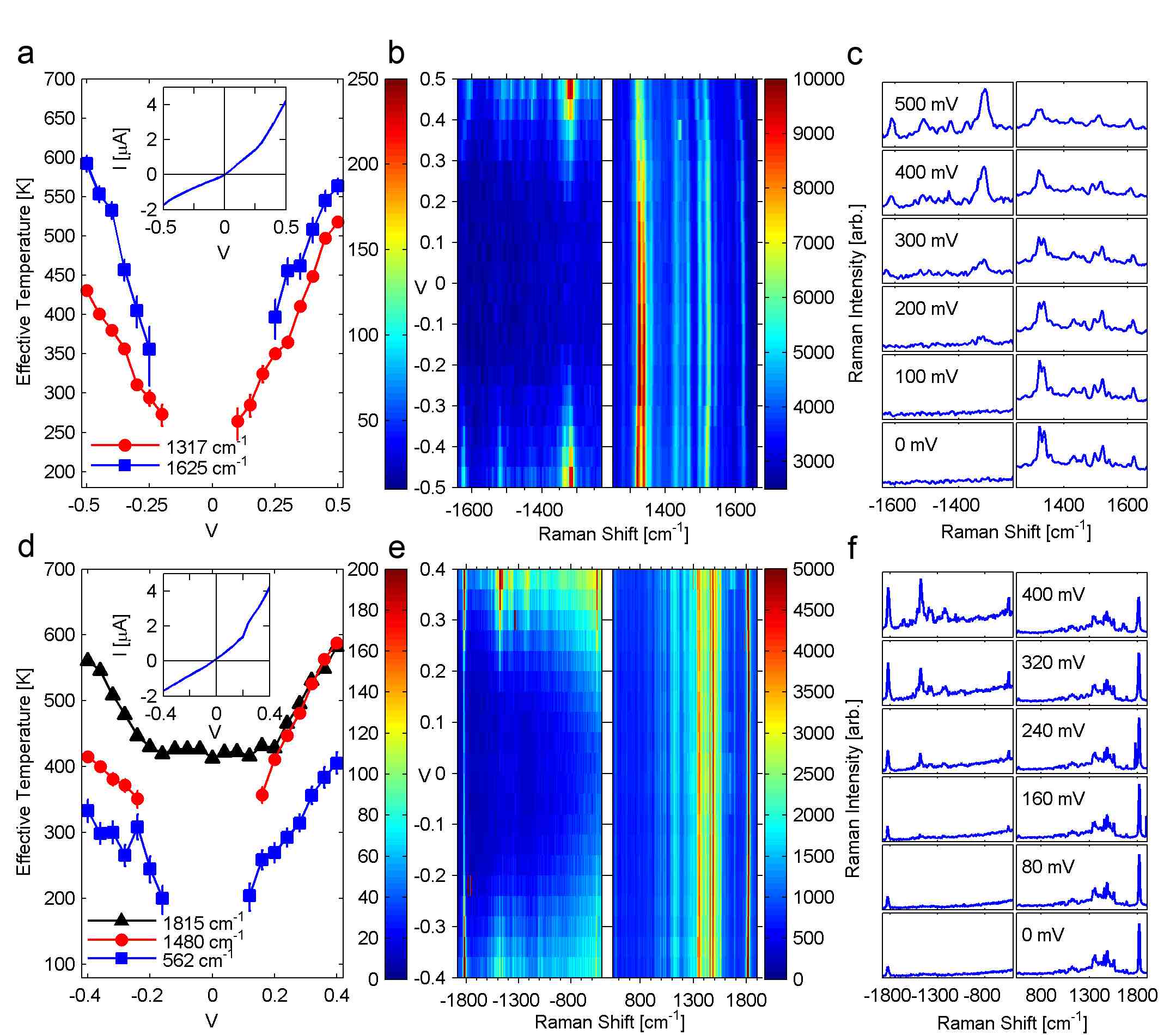}
\end{center}
\caption{{\bf Electrically driven vibrational pumping.}
a) Effective vibrational temperature as a function of $V$ for two OPV3 modes: 1317~cm$^{-1}$(red) and 1625~cm$^{-1}$(blue).  Error bars indicate uncertainty in temperature due to antiStokes amplitude measurements.
Inset) IV curve for this device.
b) Raman response of this device as a function of $V$.
c) Sample spectra for given voltage.  All Stokes and antiStokes are plotted on the same scale.  Full amplitude corresponds to 235 (10,000) counts for antiStokes (Stokes).  
d) Effective vibrational temperature as a function of $V$ for three OPV3 modes: 1815~cm$^{-1}$(black), 1480~cm$^{-1}$(red) and 562~cm$^{-1}$(blue).  Error bars indicate uncertainty in temperature due to antiStokes amplitude measurements. Simultaneous optical and electrical vibrational pumping are observed for the 1810~cm$^{-1}$ mode as T$_{eff}$ is much greater than 80~K at $V = 0$. 
e) Raman response of this devices as a function of $V$.  Blue indicates 0 counts and red indicates 200 (5000) counts for antiStokes (Stokes).
f) Sample spectra for given voltage.  All Stokes and antiStokes are plotted on the same scale.  Full amplitude corresponds to 350 (10,000) counts for antiStokes (Stokes).  
}
\end{figure}

\clearpage

\begin{figure}
\begin{center}
\includegraphics[width=6.5in]{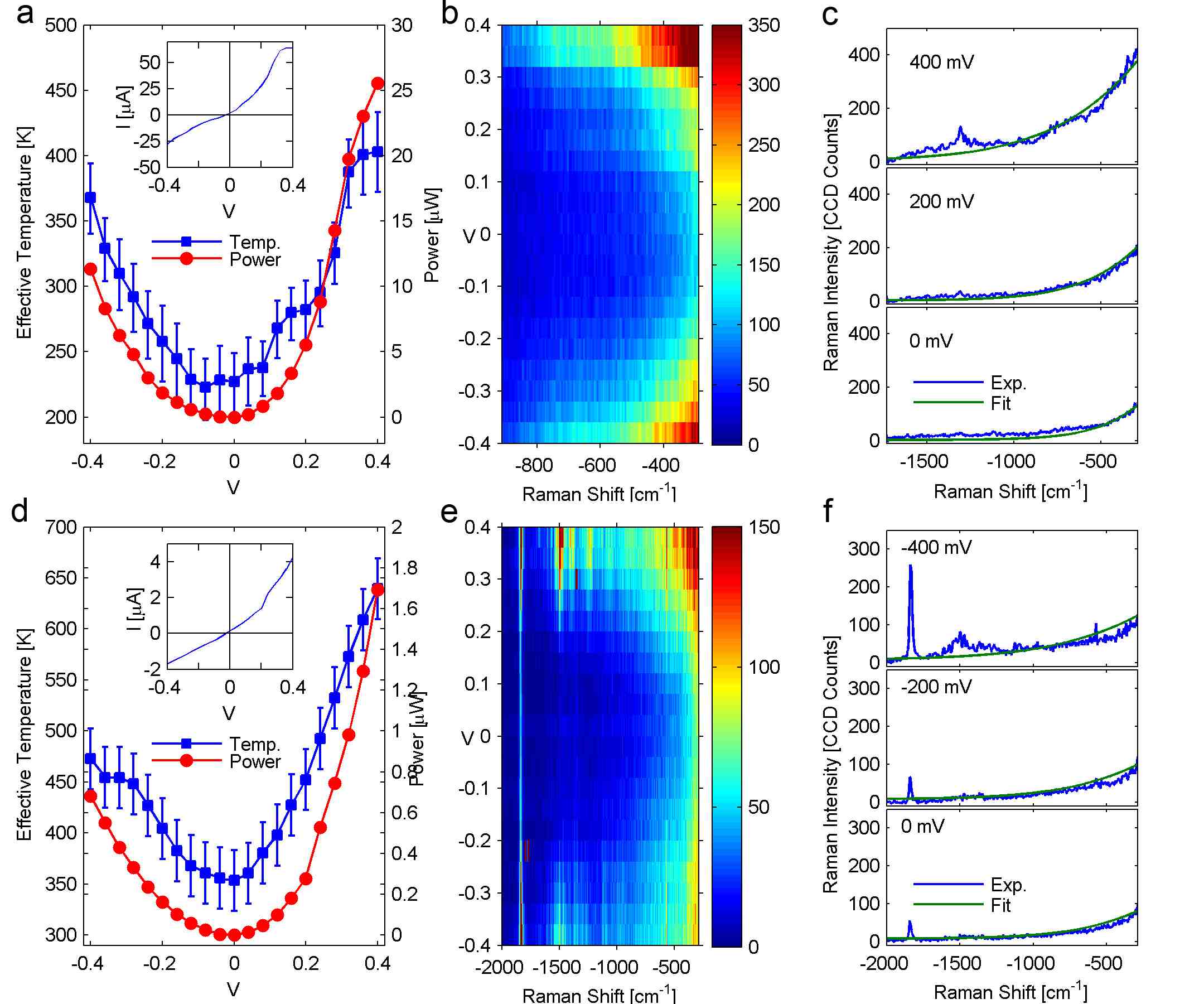}
\end{center}
\caption{{\bf Electronic heating under bias.}
a\&d) Effective temperature (blue) and dissipated electrical power (red) for two different devices.  The device in (a, b, c) shows no molecular Raman peaks, and is considered a ``clean'' junction, while (d, e, f) is the same OPV3 device that was analyzed in Figs.~1 and 3.
Insets: $I-V$ curves for these devices.  Error bars are described in text.  
b\&e) Raman response for these devices.
c\&f) Sample spectra and best fits given by Equation 2 for a given voltage.
}
\end{figure}

\clearpage


\clearpage

{\center{\bf Supporting Information: \\ Vibrational and electronic heating in nanoscale junctions}}

\renewcommand{\thefigure}{S\arabic{figure}}
\setcounter{figure}{0}

\section{Correlated electrical conductance and Raman response data}
Here are several examples of correlated electrical conductance and
Raman response from nanojunctions with OPV3 assembled on them.  In the
first three examples (Figures S1-3) the electrical conductance is on
the order of 0.01~$G_{0}$ with conductance fluctuations on the order
of 0.001~$G_{0}$.  The conductance fluctuations are correlated with
large changes in the Raman response measured at the nanojunction.
This is consistent with a single molecule switching between different
conformations in the nanojunction.  In all of these cases, we expect
that the molecule is not neatly bridging the gap, but does play a role
in total electrical conduction.  It has long been
established[S2-S4] that conduction in
such nanojunctions takes place via tunneling.  Because of the
exponential dependence of tunneling on distance, the dominant volume
for current flow is of molecular scale.  As argued
previously[S5], coincident conductance and Raman changes
imply (1) that at least some portion of a Raman-active molecule is
influencing the tunneling electrons; and (2) that Raman-active
molecule is a significant (in many cases dominant) contributor to the
total Raman signal.

At higher nanojunction conductances (0.5~$G_{0}$, such as those
observed in Figure S4), we observe similar correlations between
conduction and Raman response, but the conductance fluctuation
magnitude is much larger, as one might expect.  At this conductance
level, conduction is likely dominated by an atomic-scale metallic
junction with transmission less than one.  (One example of such a
junction would be the beginnings of a tunneling contact, with two tip
atoms slightly farther apart than their equilibrium lattice spacing.)
Again, the conduction path is highly localized, with a transverse
dimension of molecular or atomic dimensions.  Still, given the
correlation between Raman response and conductance, at least one
Raman-active molecule must be significantly coupled to this current
path.  We note that it is very unlikely that the unusually high
conductance of this particular junction results from many molecules in
parallel.  If this was the case, the observed, correlated conduction
and Raman response would imply many molecules changing their
configurations simultaneously, which is very unlikely.  One would
expect conductance fluctuations due to individual molecular motions to
be much smaller such as those seen in nanojunctions with conductances of
0.01~$G_{0}$; however this is not the case.  Instead, high
conductance nanojunctions are likely bridged by or strongly
electronically coupled to just a part of a molecule,
as the gaps are too small (based upon the conductance) to fit an
entire molecule.

\section{Junction configurations and contributions to the conductance}
Figure~\ref{fig:junctioncartoon} presents schematic examples of possible nanojunction configurations.  Fig.~\ref{fig:junctioncartoon}a shows the idealized single-molecule junction that has been considered for more than fifteen years, with a single molecule neatly bridging and bound to both ends of a nanoscale interelectrode gap.  In such a geometry, it is clear that interelectrode conduction would have a dominant contribution from current flow through the molecule.   

However, as mentioned in the main text, this ideal configuration is unlikely in electromigrated junctions, since the electromigration process (unlike mechanical break junctions or scanning tunneling microscope junctions, for example) does not control interelectrode distance with atomic precision.  Rather, the closest interelectrode distance is determined by details of individual junctions and the breaking procedure.

Fig.~\ref{fig:junctioncartoon}b-d are examples of other nanojunction geometries, in which the closest interelectrode distance is \textit{not} the length of a molecule of interest.  In Fig.~\ref{fig:junctioncartoon}b and d, only a portion of the molecule of interest may be located at the point of closest interelectrode separation, while steric effects prevent the molecule from forming strong bonds with both electrodes.  In Fig.~\ref{fig:junctioncartoon}c, the molecule is bound to both electrodes, but a closer point of interelectrode separation exists, and due to the exponential decay of tunneling with distance, the total conductance would have a dominant contribution from direct metal-metal tunneling.  These configurations, while only schematic, show clearly that many reasonable junction arrangements are possible that can have total conductances \textit{higher} than the idealized situation in Fig.~\ref{fig:junctioncartoon}a, but nonetheless involve contributions from current flow through a molecule.  An example from the literature of such a situation is thought to be presented in Tal \textit{et al.}[S6].  In that work, a benzene molecule is thought to lie down flat between two Pt electrodes in a mechanical break junction, leading to high conductance (comparable to 1 $G_{0}$), but still with molecular influence on the current. 

\section{Additional optical vibrational pumping data}
An example of optical vibrational pumping in OPV3 is shown in Figure S6. We observe several discrete spectral configurations.  Each configuration has varying levels of optical pumping, with the most noticable change occuring at 150~s, when several lower energy vibrations appear in the antiStokes signal.  

\section{Additional electrical vibrational pumping data}
An additional example of electrical vibration pumping in OPV3 is shown
in Figure S7.  The two modes observed 880~cm$^{-1}$ and 410~cm$^{-1}$
have temperature differences of almost 150~K at $V = 0.4$~V.  This
device also exhibited optical pumping of both modes resulting in zero
bias temperatures of 220~K and 120~K respectively. As expected both
modes to not show any electrical pumping until $V$ exceeds the
vibrational energy.  At 280~mV a sharp change in conductance is
observed and no pumping is observed.  The conductance changes again at
320 mV and pumping is observed but with less AS intensity than previously.  This is strong evidence the importance of the local environment to pumping cross-sections.  

\section{Additional electron heating data}
An additional example of electrical heating is shown in Figure S8.  This sample had more symmetric $I-V$ curves and power dissipation than other samples presented.  The effective temperature is observed to be more symmetric as well when considering the size of the error bars.  

\section{Raman Stark effect}
Figure S9 provides a closer look at the Raman Stark effect seen in
Figure 3B of the main text.  A clear shift of 13~cm$^{-1}$ can be
observed between zero bias and 0.5~V.  All vibrational modes present
experience similar shifts.  These shifts are reproducible on
subsequent voltage sweeps of this junction.  

\section{OPV3 Synthesis}
Tetrahydrofuran (THF) was distilled from sodium benzophenone ketyl. Triethylamine (TEA) was distilled from CaH$_2$ under N$_2$. Triethylphosphite (98\%, Aldrich), 4-aminostyrene (90\%, Aldrich), and palladium(II) acetate (98\% Aldrich)  were used as received.  Silica gel plates were 250 mm thick, 40 F254 grade obtained from EM Science. Silica gel was grade 60 (230 - 400 mesh) from EM Science.

Anhydrous DMF (120 mL) was placed in a 250 mL round bottom flask and two freeze-pump-thaw cycles were performed to ensure the removal of oxygen. A large screw-cap tube was charged with 2,5-didecoxy-1,4-diiodobenzene (10.4 g, 16.2 mmol),[S1] tetrabutylammonium bromide (15.7 g, 48.6 mmol), Pd(OAc)$_2$ (0.40 g, 1.62 mmol),  and K$_2$CO$_3$ (4.48 g, 32.4 mmol). 4-Aminostyrene (5.00 g, 32.4 mmol) was added to the DMF and the resulting solution was cannulated into the screw-cap tube. The tube was sealed with the Teflon cap and the tube was heated to 100 $^{\circ}$C for 1 d. After cooling, the crude reaction mixture was poured into water, extracted with Et$_2$O, dried over MgSO$_4$, and then the solvent was removed under reduced pressure. The crude product was then purified by silica gel chromatography using 1:1 CH$_2$Cl$_2$ : hexanes as eluent. The solid was then recrystallized from CH$_2$Cl$_2$ : hexanes to yield 1.57 g (16\%) of an orange solid. IR: 3729, 3470, 3375, 3
 207, 3025, 2925, 2851, 2648, 2321, 1976, 1622, 1599, 1516, 1466, 1413, 1254, 1172, 962, 895, 812 cm$^{-1}$. $^1$H NMR (400 MHz, CD$_2$Cl$_2$) $\delta$ 7.26 (d, J = 8.3 Hz, 4H), 7.18 (d, J = 16.5 Hz, 2 H), 7.01 (2 H), 6.96 (d, J = 16.5 Hz, 2 H), 3.95 (t, J = 6.5 Hz, 4 H), 3.73 (br s, 4 H), 1.78 (m, 4 H), 1.45 (m, 4 H), 1.34 (br m, 24 H), 0.80 (m, 6 H); $^{13}$C (100 MHz, CD$_2$Cl$_2$) $\delta$ 151.35, 147.02, 128.95, 128.91, 128.18, 127.21, 120.02, 115.51, 110.58, 70.10, 32.51, 30.26, 30.19, 30.14, 30.06, 29.96, 26.88, 23.29, 14.48. HRMS calcd for C$_{42}$H$_{60}$N$_2$O$_2$: 624.4655, found: 624.4662. 

The structure of the OPV3 Molecule is shown in Figure \ref{fig:OPV3}.

\begin{figure}
\begin{center}
\includegraphics[width=7.2in]{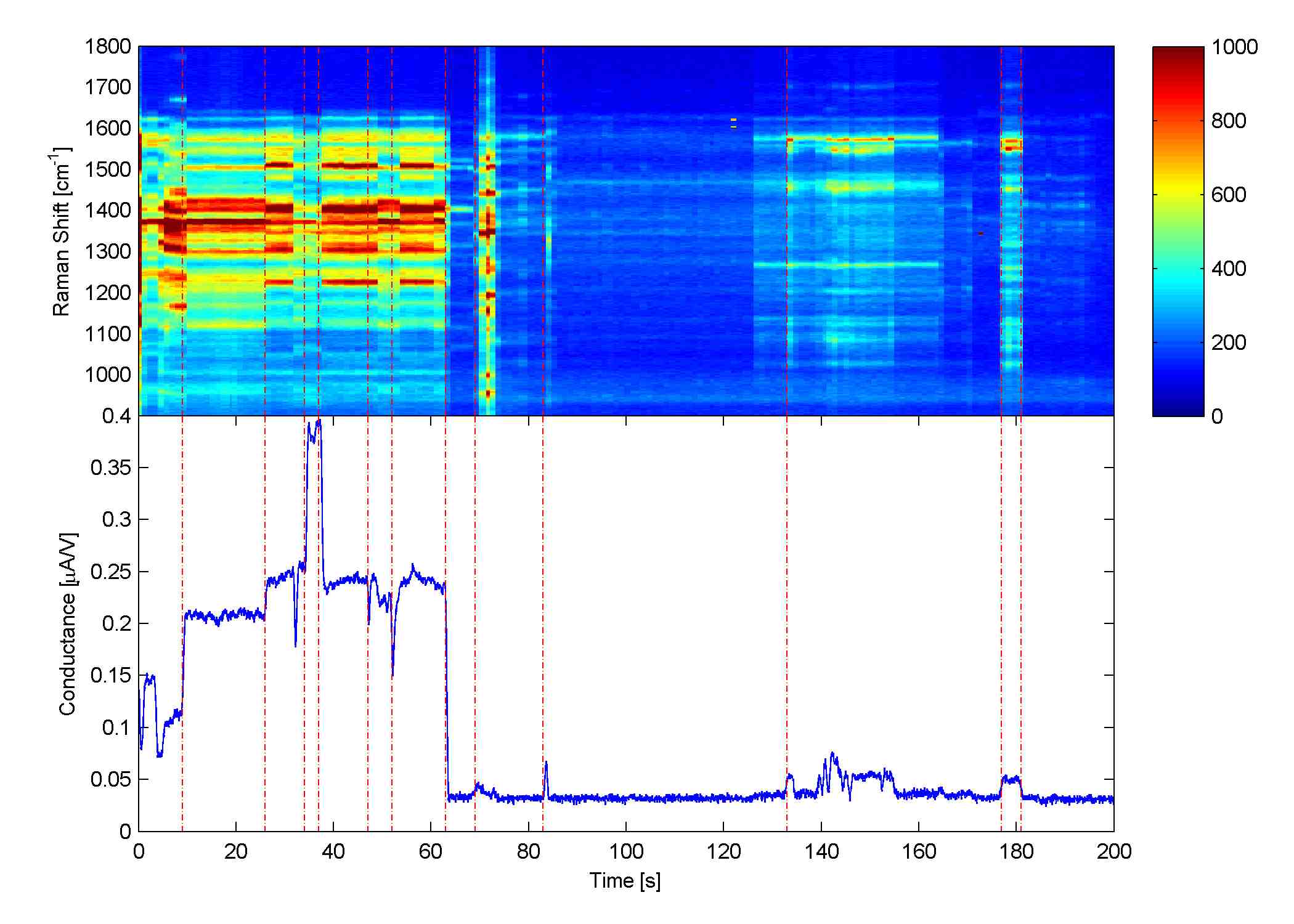}
\end{center}
\caption{
Waterfall plot of Raman spectrum (1~s integrations) and electrical conduction measurement for an OPV3 devices.  The Raman response is observed to change whenever a change in conductance occurs.  Colorbar indicates Raman intensity in CCD counts. 
}
\end{figure}

\begin{figure}
\begin{center}
\includegraphics[width=7.2in]{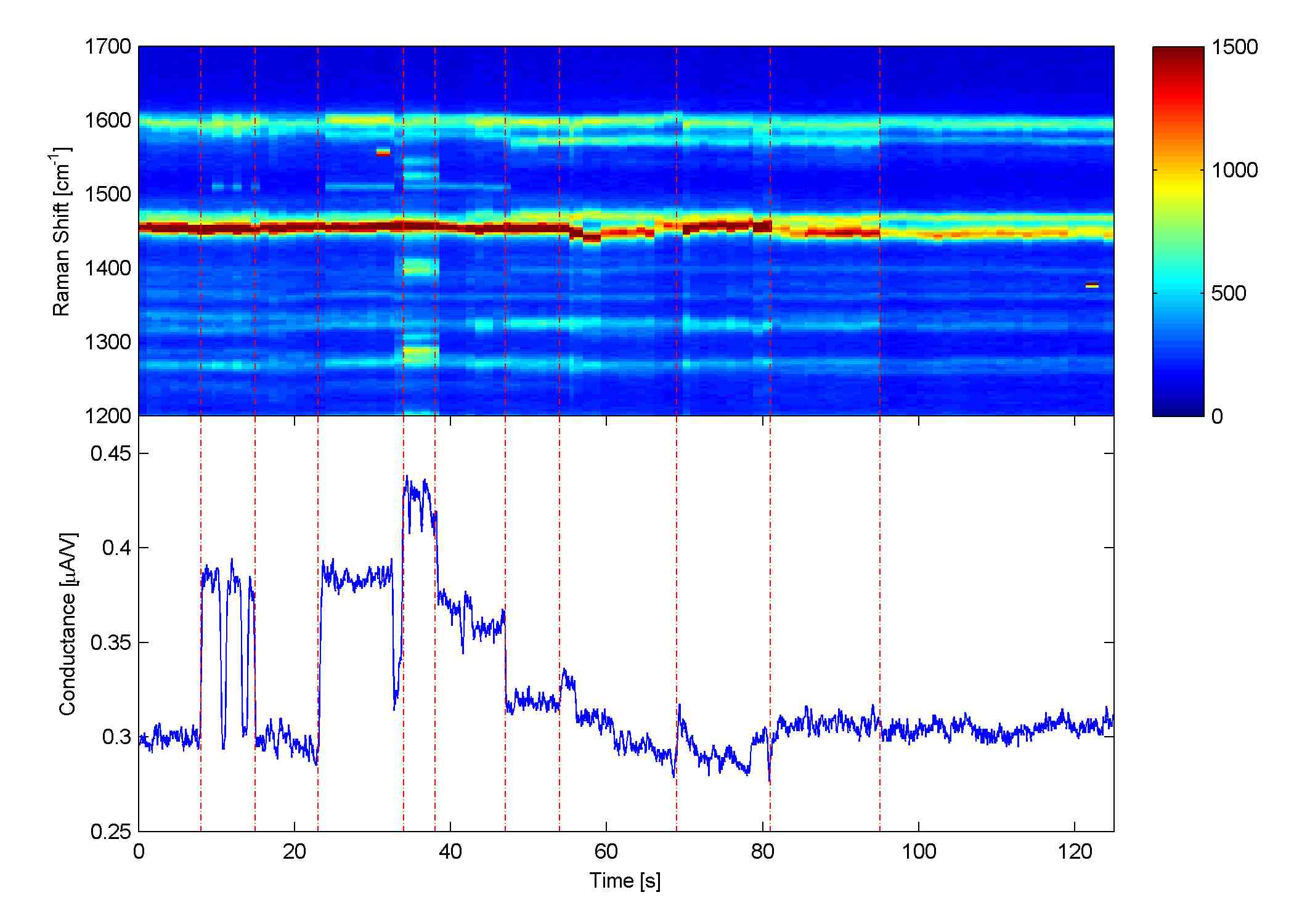}
\end{center}
\caption{
Waterfall plot of Raman spectrum (1~s integrations) and electrical conduction measurement for an OPV3 devices.  The Raman response is observed to change whenever a change in conductance occurs.  The number of Raman vibrational modes observed also changes with conductance level.  Colorbar indicates Raman intensity in CCD counts.
}
\end{figure}

\begin{figure}
\begin{center}
\includegraphics[width=7.2in]{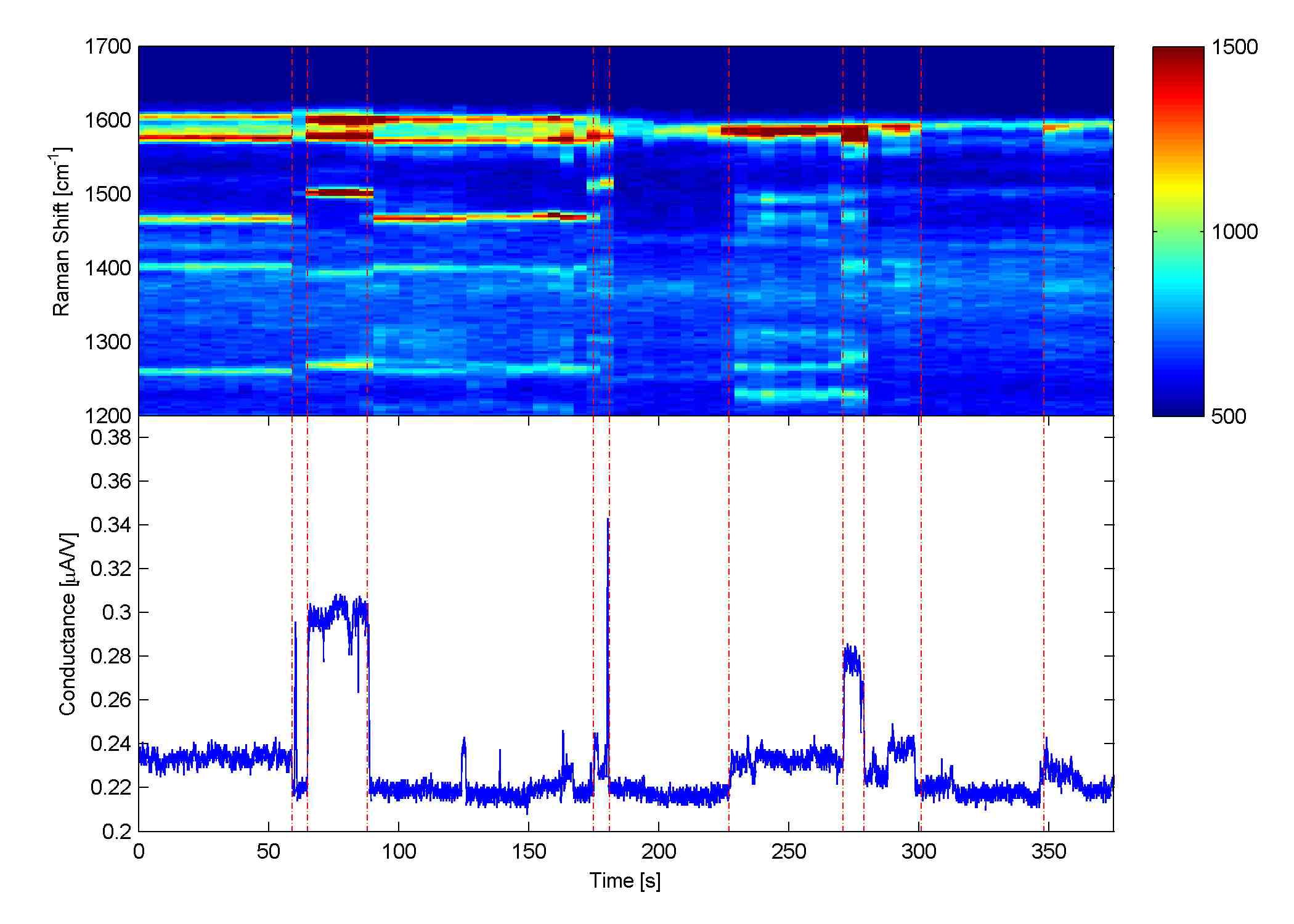}
\end{center}
\caption{
Waterfall plot of Raman spectrum (5~s integrations) and electrical conduction measurement for an OPV3 devices.  The Raman response is observed to change whenever a change in conductance occurs.  Colorbar indicates Raman intensity in CCD counts.  
}
\end{figure}

\begin{figure}
\begin{center}
\includegraphics[width=7.2in]{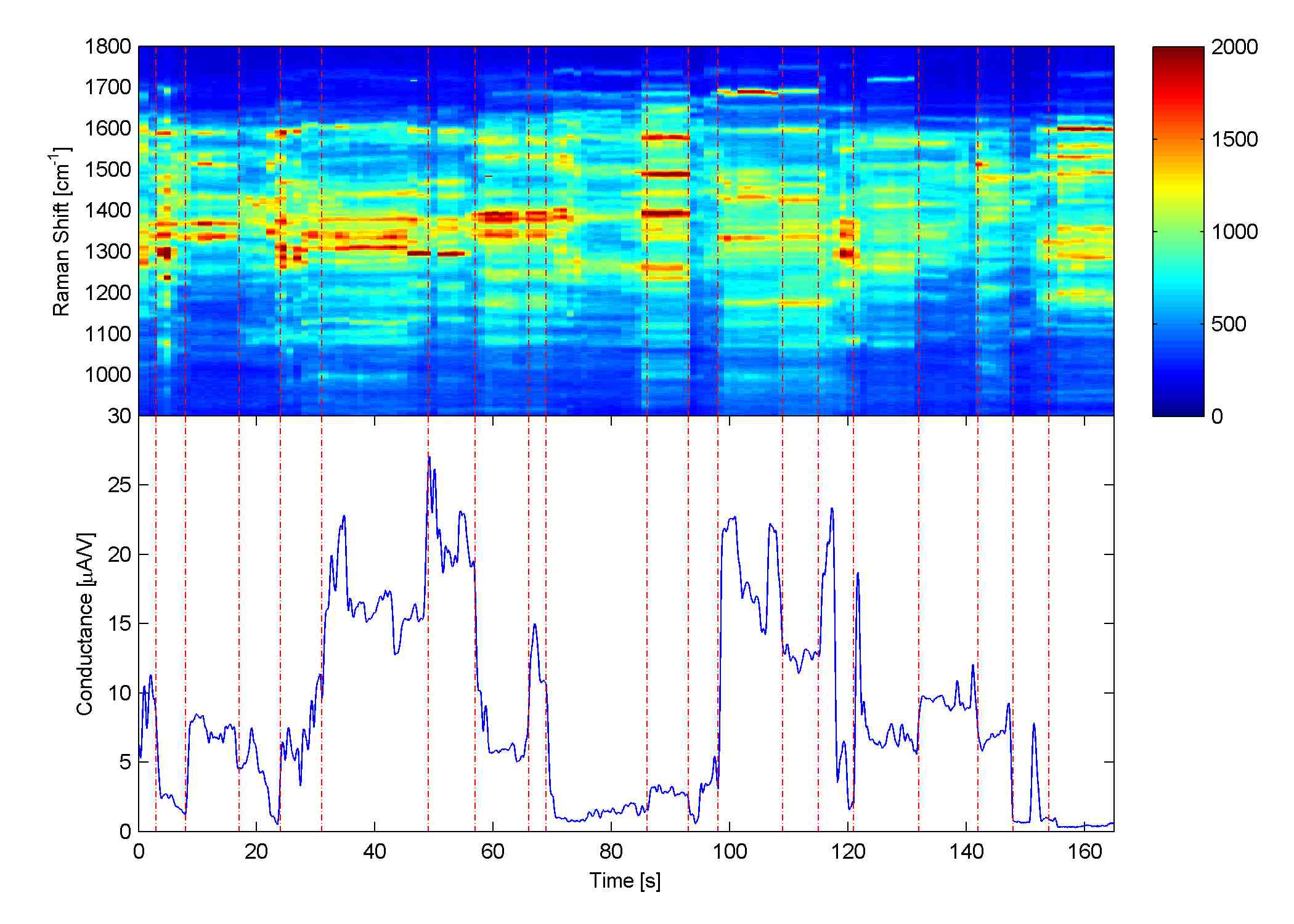}
\end{center}
\caption{
Waterfall plot of Raman spectrum (1~s integrations) and electrical conduction measurement for an OPV3 devices.  The Raman response is observed to change whenever a change in conductance occurs.  Colorbar indicates Raman intensity in CCD counts.  
}
\end{figure}

\begin{figure}
\begin{center}
\includegraphics[width=6in]{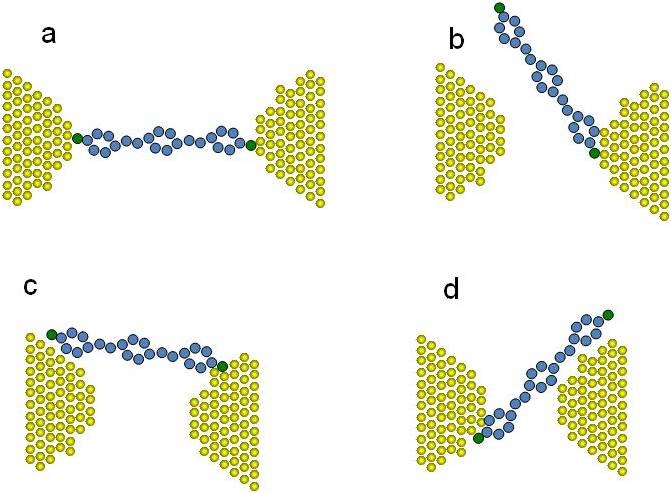}
\end{center}
\caption{\label{fig:junctioncartoon}
Cartoon of possible junction configurations, using the OPV3 molecule as an example (saturated side chains omitted for clarity).  (a) The idealized single-molecule junction, a configuration unlikely in these experiments since interelectrode separation is not precisely controlled.  (b-d) Alternative junction configurations, in which interelectrode conduction would be expected to include a contribution from current interacting with the molecule, as well as direct metal-metal tunneling.
}
\end{figure}

\begin{figure}
\begin{center}
\includegraphics[width=7.0in]{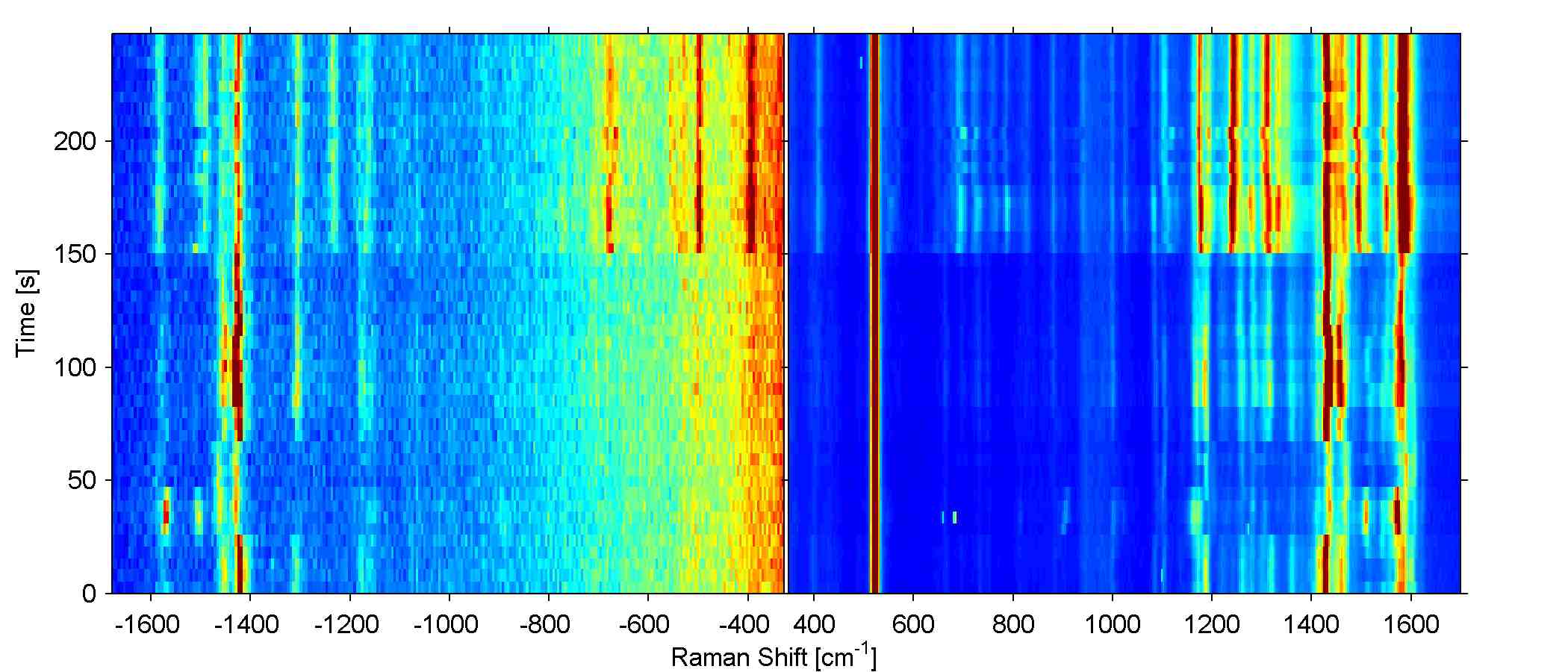}
\end{center}
\caption{
Raman response of an OPV3 junction as a function of time under zero bias.   Blue indicates 0 counts and red indicates 100 (8000) counts for antiStokes (Stokes) sides.  Integration time is 1~s.  The junction switches stochastically between several stable configurations, each with characteristic spectra that exhibit strong optical pumping of different vibrational modes.
}
\end{figure}

\begin{figure}
\begin{center}
\includegraphics[width=7.2in]{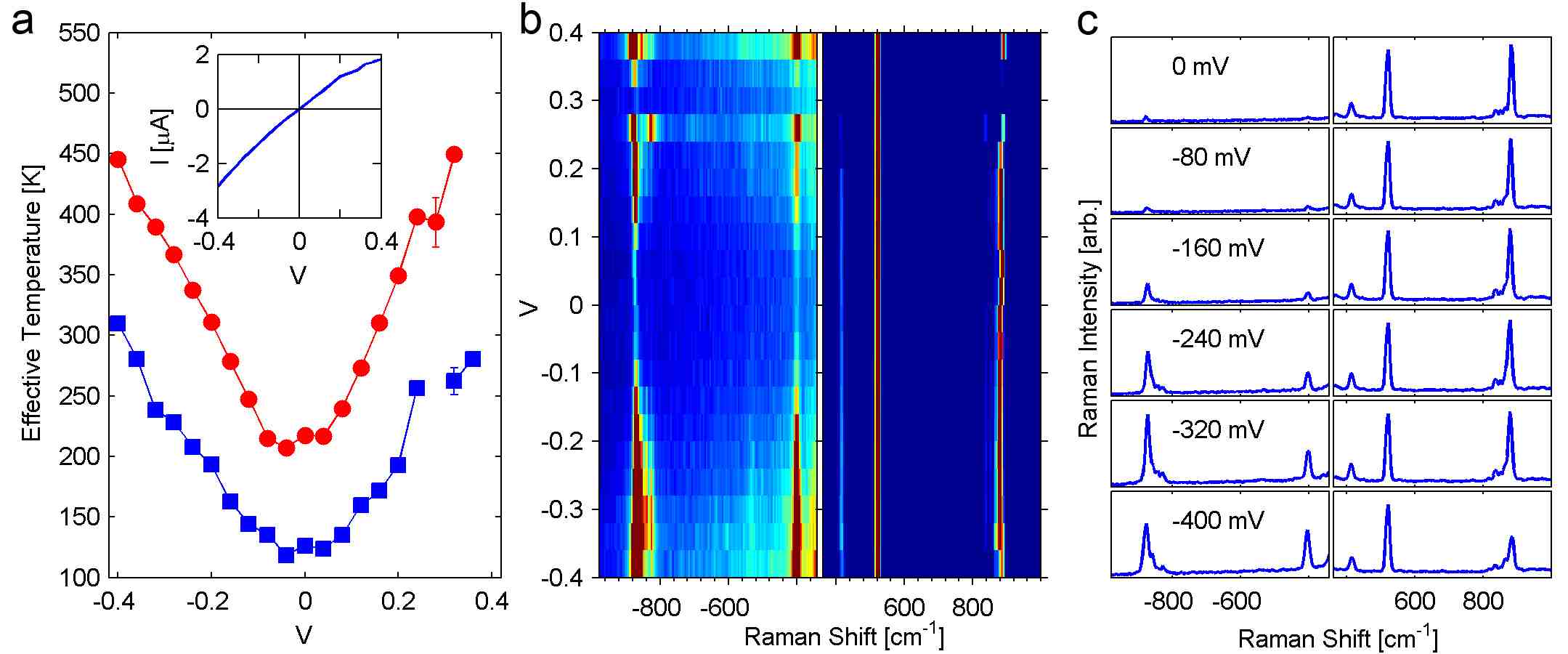}
\end{center}
\caption{
A) Effective vibrational temperature as a function of $V$ for two OPV3
modes: 880~cm$^{-1}$(red) and 410~cm$^{-1}$(blue).  Error bars
indicate the uncertainty in inferred effective temperature due to the
statistical limitations of the antiStokes amplitude
measurements.
Inset) IV curve for this device.
B) Raman response of this device as a function of $V$.  Blue indicates
10 (2500) counts and red indicates 250 (10,000) counts for antiStokes
(Stokes).  The strong Stokes peak at 520~cm$^{-1}$ is from the Si substrate.
C) Sample spectra for given voltage.  All antiStokes (Stokes) spectra are plotted on the same scale.  Full amplitude corresponds to 1200 (18,000) counts for antiStokes (Stokes).  
}
\end{figure}

\begin{figure}
\begin{center}
\includegraphics[width=7.2in]{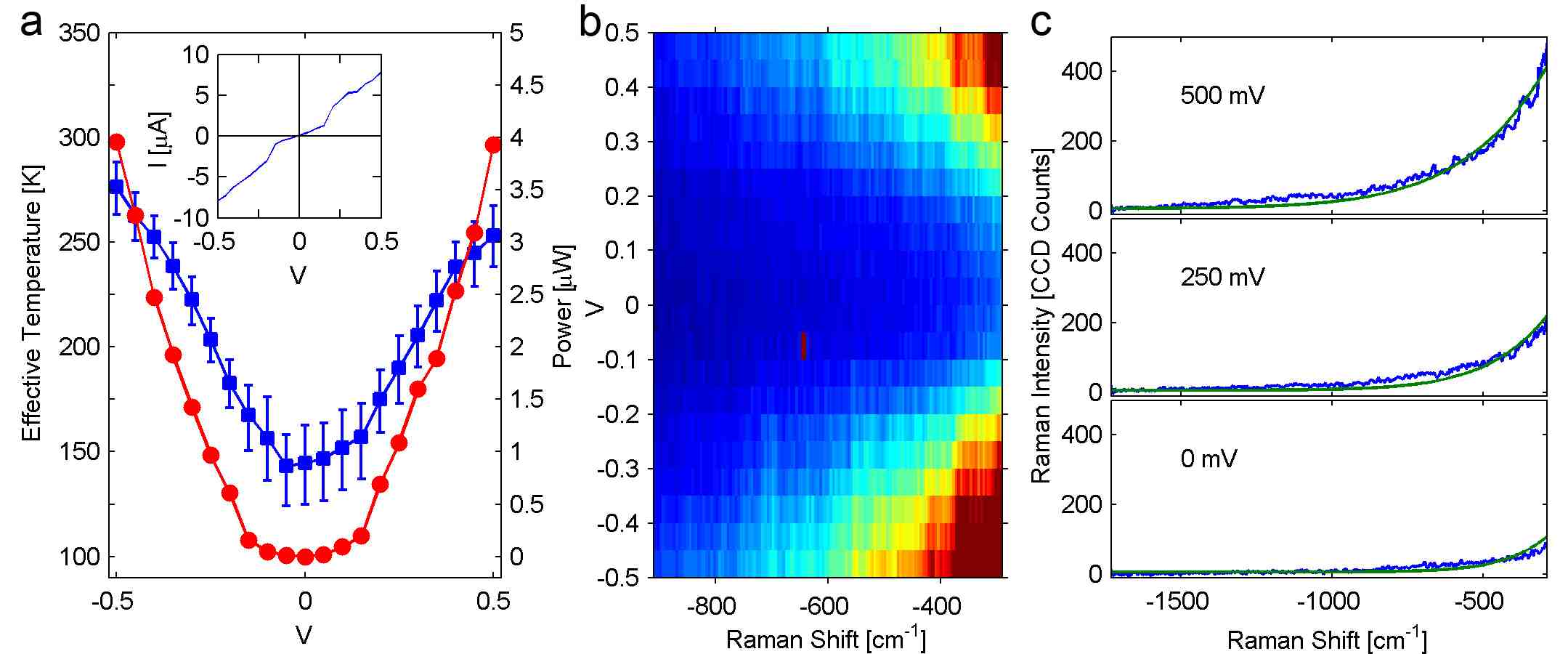}
\end{center}
\caption{
A)Effective electronic temperature(blue) and dissipated electrical power (red) for a device with very little vibrational Raman activity.  Error bars are described in text.  
Inset) $I-V$ curves for these devices. 
B) Raman response for these devices.  Blue indicates 0 counts and red indicates 350 counts.  
C) Sample spectra (blue) and best fit given by Equation 2 in main text (green) for a given voltage.
}
\end{figure}

\begin{figure}
\begin{center}
\includegraphics[width=7.2in]{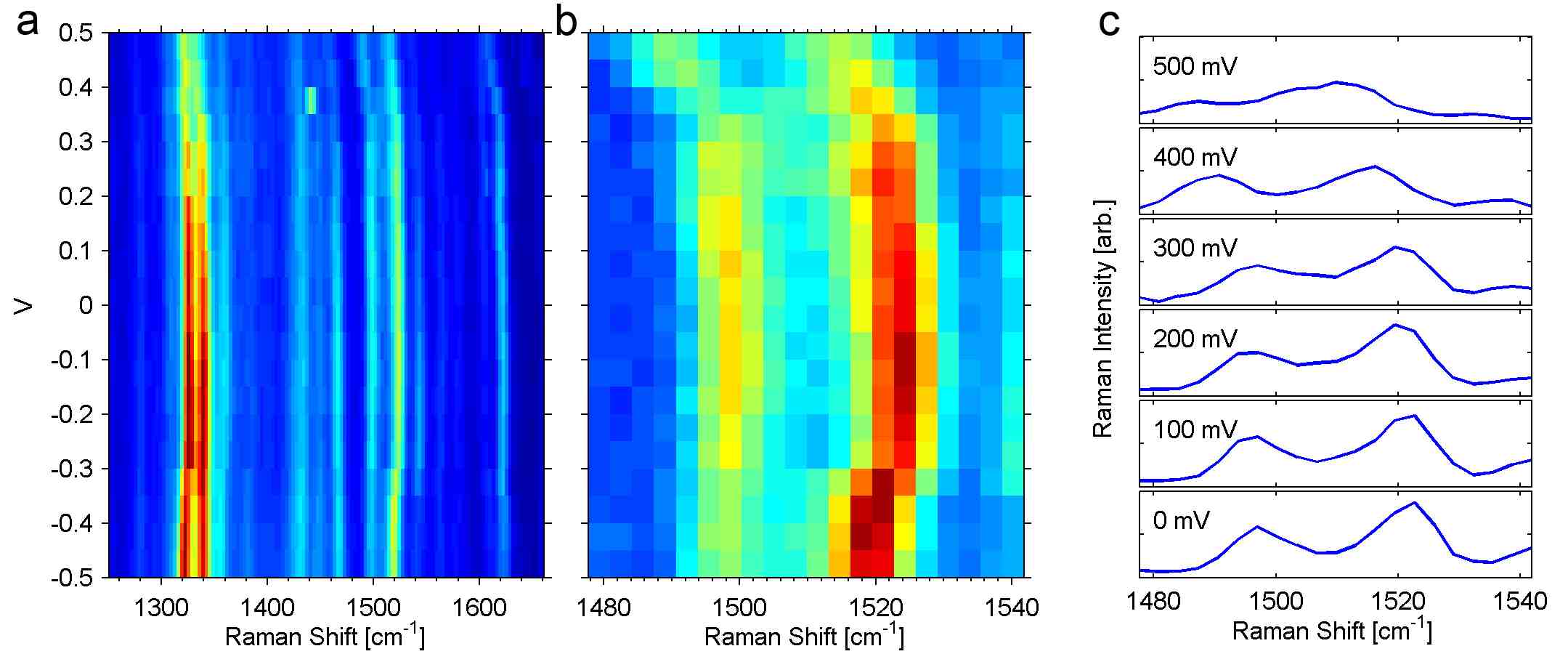}
\end{center}
\caption{
A)  Raman response of the same device shown in Figure 3a in main text.  A shift of about 15~cm$^{-1}$ is present for many spectral lines. Blue indicates 2500 counts and red indicates 10000 counts.   
B) Zoom in of Raman response for the mode centered at 1523~cm$^{-1}$.Blue indicates 2500 counts and red indicates 7000 counts.   
C) Sample spectra for given voltage. All spectra are plotted on the same scale with a base line of 3000 counts substracted.  Full amplitude corresponds to 4000 counts.  The peak at 1523~cm$^{-1}$ can clearly be seen systematically shifting to lower energy at higher voltages reaching 1510~cm$^{-1}$ at 500~mV.
}
\end{figure}

\begin{figure}
\begin{center}
\includegraphics[width=4in]{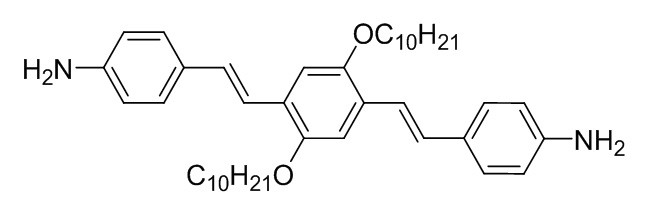}
\end{center}
\caption{\label{fig:OPV3}
OPV3 Molecule
}
\end{figure}

\clearpage

\noindent[S1]  Zhao, Y., Shirai, Y., Slepkov, A.~D., Cheng, L., Alemany, L.~B.,
Sasaki, T., Hegmann, F.~A., Tour, J.~M. Synthesis, Spectroscopic and
Nonlinear Optical Properties of Multiple [60]Fullerene-Oligo(
\textit{p}-Phenylene Ethynylene) Hybrids.  {\it Chem. Eur. J.} {\bf 11}, 3643
(2005).

\noindent[S2]  Park, H. et al.  Fabrication of metallic electrodes with nanometer separation by electromigration.  {\it Appl. Phys. Lett.} {\bf 75}, 301-303 (1999).

\noindent[S3]  Natelson, D., Yu,  L. H., Ciszek, J. W., Keane, Z. K. \& Tour, J. M. Single-molecule transistors: electron transfer in the solid state.  {\it Chem. Phys.} {\bf 324}, 267-275 (2006).

\noindent[S4]  Ward, D.~R., Scott, G.~D., Keane, Z.~K., Halas, N.~J. \& Natelson, D.  Electronic and optical properties of electromigrated molecular junctions.  {\it J. Phys. Condens. Matt.} {\bf 20}, 374118 (2008).

\noindent[S5]  Ward, D.~R. et al.  Simultaneous measurements of electronic conduction and Raman response in molecular junctions.  {\it Nano Lett.\/} {\bf 8}, 919-924 (2008).

\noindent[S6]  Tal, O. et al.  Molecular signature of highly conductive metal-molecule-metal junctions.  {\it Phys. Rev. B} {\bf 80}, 085427 (2009).

\end{document}